\documentclass[12pt]{article}
\pdfoutput=1

\usepackage{amsmath}
\usepackage{amsfonts}
\usepackage{amssymb}
\usepackage{dsfont}

\usepackage{hyperref}
\hypersetup{colorlinks=true, linkcolor=red, citecolor=blue, linktoc=page}

\def\cA{{\cal A}}
\def\cB{{\cal B}}

\def\cG{{\cal G}}

\def\cN{{\cal N}}

\def\Re{{\rm Re \,}}
\def\Im{{\rm Im \,}}

\def\p{\partial}

\usepackage{color}
\usepackage{graphicx}

\usepackage{sectsty}
\sectionfont{\large}

\renewcommand{\thefootnote}{\fnsymbol{footnote}}
\newcommand{\starttext}{
\setcounter{footnote}{0}
\renewcommand{\thefootnote}{\arabic{footnote}}}
\newcommand{\bea}{\begin{eqnarray}}
\newcommand{\eea}{\end{eqnarray}}
\newcommand{\be}{\begin{equation}}
\newcommand{\ee}{\end{equation}}

\newcommand{\no}{\nonumber}

\numberwithin{equation}{section}

\renewcommand{\Im}{\operatorname{Im}}
\renewcommand{\Re}{\operatorname{Re}}

\DeclareMathOperator{\Vol}{Vol}

\long\def\symbolfootnote[#1]#2{\begingroup%
\def\thefootnote{\fnsymbol{footnote}}\footnote[#1]{#2}\endgroup}

\begin{document}
\setlength{\baselineskip}{16pt}

\starttext
\setcounter{footnote}{0}

\begin{flushright}
IFT-UAM/CSIC-18-052 \\
\today
\end{flushright}

\bigskip

\begin{center}

{\Large \bf   Massive spin 2 excitations \\[5pt] in $AdS_6\times S^2$ warped spacetimes}

\vskip 0.4in

{\large   Michael Gutperle$^a$, Christoph F.~Uhlemann$^a$ and Oscar Varela$^{b,c,d}$}

\vskip 0.2in

{ \sl $^a$ Mani L. Bhaumik Institute for Theoretical Physics} \\
{\sl Department of Physics and Astronomy }\\
{\sl University of California, Los Angeles, CA 90095, USA} 

\vskip 0.2in

{ \sl $^b$ Department of Physics, Utah State University, Logan, UT 84322, USA} \\

\vskip 0.2in

{ \sl $^c$ Departmento de F\'\i sica Te\'orica and Instituto de F\'\i sica Te\'orica UAM/CSIC , }\\ {\sl  Universidad Aut\'onoma de Madrid, Cantoblanco, 28049 Madrid, Spain} \\

\vskip 0.2in

{ \sl $^d$  Max-Planck-Institut f\"ur Gravitationsphysik (Albert-Einstein-Institut), }\\ {\sl  Am M\"uhlenberg 1, D-14476 Potsdam, Germany } \\

\bigskip

\end{center}
 
\begin{abstract}
\setlength{\baselineskip}{18pt}

\noindent We study (massive) spin-2 fluctuations around warped $AdS_6$ solutions with $16$ supersymmetries in type IIB supergravity. We identify two classes of fluctuations which are universally present for all solutions of this form. The holographically dual operators have scaling dimensions $\Delta=5+3\ell$ and $\Delta=6+3\ell$, where the integer $\ell$ encodes the $R$-symmetry charge. They are identified as descendant states in respective BPS multiplets (the current multiplet for $\Delta=5$). We also compute the normalization of the energy-momentum tensor two-point function and show that it is related to the $S^5$ partition function of the dual $5d$ SCFTs.

\end{abstract}

\setcounter{equation}{0}
\setcounter{footnote}{0}

\newpage

\section{Introduction}

Supersymmetric conformal field theories  (SCFT)  in five dimensions exhibit many interesting properties: They are intrinsically strongly coupled in the UV and many have relevant deformations that flow to non-renormalizable gauge theories in the IR. Generically, they do not have exactly marginal deformations and no conventional Lagrangian description.  These features are challenging for (perturbative) QFT methods, but the existence and properties  of such theories can be studied using string theory and  M-theory \cite{Seiberg:1996bd,Morrison:1996xf,Intriligator:1997pq}.   

AdS/CFT is a particularly powerful tool to quantitatively study these theories, and supergravity solutions describing five dimensional SCFTs have been found in type IIA \cite{Brandhuber:1999np,Bergman:2012kr,Passias:2012vp} and type IIB supergravity \cite{DHoker:2016ujz,DHoker:2016ysh,DHoker:2017mds}\footnote{See \cite{Lozano:2012au,Lozano:2013oma,Apruzzi:2014qva,Kelekci:2014ima,Kim:2015hya,Kim:2016rhs} for earlier work on type IIB solutions.}. The type IIB solutions were identified with a large class of  SCFTs which are obtained from the conformal limit of $(p,q)$ 5-brane webs \cite{Aharony:1997bh,Aharony:1997ju}. The IIB  supergravity solutions have since been studied and extended. In \cite{Gutperle:2017tjo} the $S^5$ partition function and entanglement entropy for a spherical region where calculated holographically. In \cite{Kaidi:2017bmd} probe $(p,q)$ strings in the supergravity background  have been studied. In \cite{DHoker:2017zwj} supergravity solutions were constructed with punctures and non-trivial $SL(2,\mathds{R})$ monodromy, and they were related in a precise way to 5-brane webs with additional 7-branes in \cite{Gutperle:2018vdd}.

An important part of the basic data of any SCFT is the spectrum of short and long operators and how they fit into representations of the superconformal algebra. It is well known that five dimensional superconformal symmetry is a special case, with a unique superconformal algebra $F(4)$  \cite{Nahm:1977tg,Kac:1977em,Shnider:1988wh}. Recently, the superconformal multiplets of $F(4)$  have been classified in \cite{Buican:2016hpb,Cordova:2016emh}. On general grounds, it is expected that linearized  fluctuations of the supergravity fields around an $AdS$  vacuum fall into superconformal multiplets, with  the complete spectrum determined by the Kaluza-Klein (KK) reduction. A famous example is the KK-reduction of type IIB supergravity on its $AdS_5\times S^5$ vaccum  \cite{Kim:1985ez,Gunaydin:1984fk}
and the relation to the spectrum of  half-BPS operators of $\cN=4$ SYM. The $AdS_6$ solutions of type IIB supergravity are, however, more complicated: they only preserve half the number of supersymmetries, the metric is a more involved warped product of $AdS_6\times S^2$ over a Riemann surface, and they involve non-trivial configurations of the axion-dilaton scalar and complex two-form field. A general analysis of the coupled equations of motion and the resulting KK spectrum for these warped compactifications is therefore challenging, and we instead focus on a special class of fluctuations which can be studied by itself.

The structure of the present paper is as follows: In  section \ref{sec2} we briefly review the type IIB solutions of \cite{DHoker:2016ujz,DHoker:2016ysh,DHoker:2017mds,DHoker:2017zwj} and their relation to 5-brane webs. In section \ref{sec3} we study fluctuations of the metric field which correspond to massless or  massive  symmetric, transverse  and traceless rank two tensors on $AdS_6$. Crucially, the linearized equation of motion for these fluctuations decouples from all other supergravity fields \cite{Bachas:2011xa}, and we identify a particular class of solutions which is universally present for the type IIB $AdS_6$ solutions discussed above. As we discuss in section \ref{sec-scft}, in the dual SCFTs these fluctuations source the stress tensor (in the massless case) as well as spin two tensor operators of higher scaling dimension. We explicitly obtain the ``central charge" $C_T$ which appears in the two point function of the stress tensor from our supergravity fluctuation. We close with a discussion in section \ref{sec4}.

\section{Type IIB \texorpdfstring{$AdS_6\times S^2$}{AdS6xS2} warped solution}\label{sec2}

In this section we review aspects of the supergravity solutions constructed in \cite{DHoker:2016ujz} which will be important in the present paper. These solutions provide the general local form of warped $AdS_6$ solutions with $16$ supersymmetries in type IIB supergravity. The geometry is a warped product of $AdS_6\times S^2$ over a Riemann surface $\Sigma$. The complex scalar and the complex two-form potential (proportional to the volume form on $S^2$) depend on the coordinates on $\Sigma$. The metric is given by 
\be
\label{2a1}
ds^2  =   f_6^2 \, ds^2 _{AdS_6} + f_2 ^2 \, ds^2 _{S^2} +  4 \rho^2 | dw |^2~.
\ee
The metric functions and the remaining supergravity fields are all expressed in terms of two locally holomorphic functions ${\cal A}_\pm$ on $\Sigma$, via the composite quantities
\begin{align}\label{eq:composites}
 \kappa ^2 & =  - |\partial_w \cA_+|^2 + |\partial_w \cA_-|^2~, & 
 \p_w \cB &= \cA_+ \p_w \cA_- - \cA_- \p_w \cA_+~,
 \no\\
 \cG & =  |\cA_+|^2 - |\cA_-|^2 + \cB + \bar \cB~,&
 R+\frac{1}{R} &=  2+ { 6 \, \kappa^2 \, \cG \over |\partial_w\cG|^2}~.
\end{align}
Note that $\cG$ and $\kappa^2$ are related as follows
\bea
\label{2b3}
\p_{\bar w} \p_w \cG = - \kappa ^2~.
\eea
In Einstein frame, the metric functions $f_6^2$, $f_2^2$ and $\rho^2$ are given by
\bea\label{2a4}
f_6^2=\sqrt{6\cG} \left ( \frac{1+R}{1-R} \right ) ^{\tfrac{1}{2}},
\hskip 0.4in
f_2^2=\frac{1}{9}\sqrt{6\cG} \left ( \frac{1-R}{1+R} \right )^{\tfrac{3}{2}},
\hskip 0.4in
\rho^2={\kappa^2 \over \sqrt{6\cG} } \left (\frac{1+R}{1-R} \right )^{\tfrac{1}{2}}.
\eea
The expressions for the axion-dilaton scalar and the complex two-form $C_{(2)}$ can be found in \cite{DHoker:2016ujz}, but are not needed in the following. For arbitrary locally holomorphic $\cA_\pm$ this yields a solution to the BPS equations (and also to the equations of motion \cite{Corbino:2017tfl}). For physically regular solutions, $\kappa^2$ and $\cG$ satisfy
\begin{align}\label{eq:reg1}
 \kappa^2&>0~, & \cG&>0~,
\end{align}
at interior points of $\Sigma$. If $\Sigma$ has a boundary, they further satisfy the boundary conditions
\begin{align}\label{eq:reg2}
 \kappa^2\big\vert_\Sigma&=\cG\big\vert_\Sigma=0~,
\end{align}
which ensure that the $S^2$ collapses on $\partial\Sigma$ to smoothly close off the $10d$ geometry.
Large classes of physically regular solutions were constructed in \cite{DHoker:2016ysh,DHoker:2017mds,DHoker:2017zwj}, and related to $(p,q)$ 5-brane webs in type IIB string theory. The Riemann surface $\Sigma$ is a disc in these solutions, and the differentials $\partial_w\cA_\pm$ have poles at isolated points on the boundary of the disc. The technical details of these constructions will be largely irrelevant for the discussion in the present paper. The results hold in particular for these solutions, but are not limited to them. We will make use of the regularity conditions (\ref{eq:reg1}) and (\ref{eq:reg2}), and discuss the poles where necessary, e.g.\ in relation to regularity conditions in sec.~\ref{sec:bc} and \ref{sec33}.

\section{The spectrum of spin 2 excitations}\label{sec3}

As discussed in the introduction, determining the complete KK-spectrum of a warped $AdS$ supergravity solution is challenging. However, as shown in \cite{Bachas:2011xa}, the spectrum of spin-$2$ fluctuations is determined from a universal equation which depends on the geometry only, not on the other fields in type IIB supergravity.  The original motivation of  \cite{Bachas:2011xa} was to study localized spin 2 excitations on solutions dual to defect conformal field theories. The result on the decoupling, however, can be applied to general warped spacetimes with $AdS$ factors, see e.g. \cite{Klebanov:2009kp,Richard:2014qsa,Passias:2016fkm,Schmude:2016bqp,Pang:2017omp,Passias:2018swc} for application in various cases.

While the formulae in \cite{Bachas:2011xa} were for $AdS_4$, it is straighforward to generalize the analysis to the case which we are interested in, namely the  fluctuations around the warped $AdS_6\times S^2\times\Sigma$ solutions. 
The perturbed $10d$ metric takes the form
\begin{align}\label{eq:delta-ds2}
 ds^2&=f_6^2\left(ds^2_{AdS_6}+h_{\mu\nu}dx^\mu dx^\nu\right)+\hat g_{ab}dy^a dy^b~,
\end{align}
with $h_{\mu\nu}$ a transverse traceless fluctuation in the unit-radius $AdS_6$ part of the background geometry,
and the metric on the 4-dimensional internal space in (\ref{2a1}) given by
\begin{align}\label{eq:int-metric}
 \hat g_{ab}dy^a dy^b &= f_2^2 ds^2_{S^2}+ 4 \rho^2 |dw|^2~.
\end{align}
The $6d$ part of the metric perturbation is expanded in modes on unit-radius $AdS_6$,
\begin{align}\label{eq:h-modes}
 h_{\mu\nu}(x,y)&=h^{[tt]}_{\mu\nu}(x) \psi(y)~,
 &
 \square_{AdS_6}h^{[tt]}_{\mu\nu}&= ( M^2 -2) h^{[tt]}_{\mu\nu}~.
\end{align}
The starting point is eq.~(2.23) of \cite{Bachas:2011xa}, which holds for a non-compact part of arbitrary dimension.
With $g$ denoting the full $10d$ metric and $M,N=0,..,9$, it is given by
\begin{align}\label{eq:Laplace0}
 \frac{1}{\sqrt{-g}}\partial_M\sqrt{-g} g^{MN}\partial_N h_{\mu\nu}&=0~.
\end{align}
With the mode expansion (\ref{eq:h-modes}), this reduces to an equation for $\psi$, given by
\begin{align}\label{laplace1}
-\frac{1}{f_6^{4}\sqrt{\hat g}} \partial_a ( f_6^6\sqrt{\hat g} \hat g^{ab}  \partial_b) \psi=M^2 \psi~.
\end{align}
The only effect of having $AdS_6$ instead of $AdS_4$ are the different powers of the $AdS$ warp factor as compared to \cite{Bachas:2011xa}.
With the explicit form of the internal space metric in (\ref{eq:int-metric}),\footnote{The $\Sigma$ part of the metric in (\ref{2a1}) may be written more explicitly as $g_\Sigma=2\rho^2(dw\otimes d\bar w+d\bar w\otimes dw)$, from which the components of the metric and inverse metric are extracted.} eq.~(\ref{laplace1}) evaluates to
\begin{align}\label{eq:psi}
{1\over f_6^4 f_2^2 \rho^2} \partial_a\big(f_6^6f_2^2  \eta^{ab} \partial_b \psi\big )+ {f_6^2 \over f_2^2} \nabla_{S^2}^2 \psi + M^2 \psi&=0~,
\end{align}
where $a,b=w,\bar w$ from now on and 
\begin{align}\label{eq:eta}
 \eta^{w\bar w}&=\eta^{\bar w w}=\frac{1}{2}~.
\end{align} 
Somewhat remarkably, the composite quantity $R$ cancels in all the combinations of metric factors, which are given explicitly by
\begin{align}\label{eq:metric-combinations}
 {f_6^2\over f_2^2} &= 9+ {6|\partial \cG|^2 \over \kappa^2\cG }~,
 &
f_6^4f_2^2\rho^2 &=\frac{2}{3}\kappa^2\cG~,
&
f_2 f_6^3 &= 2 \cG~.
\end{align}
We further expand $\psi$ in spherical harmonics on $S^2$
\begin{align}\label{eq:psi-exp}
 \psi(y)&=\phi_{\ell m}(w,\bar w) Y_{\ell m}(S^2)~.
\end{align}
This turns eq.~(\ref{eq:psi}) into
\begin{align}\label{eq:phi-eq}
 6 \partial_a\big(\cG^2  \eta^{ab} \partial_b \phi_{\ell m}\big )- \ell(\ell+1)\left(9\kappa^2\cG+6|\partial\cG|^2\right) \phi_{\ell m} + M^2\kappa^2\cG  \phi_{\ell m}&=0~.
\end{align}
This equation can be further simplified with the field redefinition
\begin{align}\label{eq:chi-def}
 \phi_{\ell m}&=\cG^\ell \chi_{\ell m}~.
\end{align}
After multiplying (\ref{eq:phi-eq}) by $\cG^{\ell}$, it becomes
\begin{align}\label{eq:chi-eq}
 \partial_a \left(\cG^{2\ell+2}\partial^a\chi_{\ell m}\right)+\frac{1}{6}\left(M^2-3\ell(3\ell+5)\right)\kappa^2\cG^{2\ell+1}\chi_{\ell m}&=0~.
\end{align}
This equation determines the transverse traceless spin-2 fluctuations around generic type IIB warped $AdS_6\times S^2\times\Sigma$ solutions reviewed in sec.~\ref{sec2}.

\subsection{Regularity conditions}\label{sec:bc}

We implement the following regularity requirements for the metric fluctuations. In the interior of $\Sigma$, the perturbation should be finite. At generic points on the boundary of $\Sigma$, $\partial\Sigma$, the $S^2$ closes off, as ensured by the boundary condition (\ref{eq:reg2}). These boundary points are completely regular points in the $10d$ geometry, and as such the metric fluctuation at these points should be finite. Locally, coordinates can be chosen such that $\Sigma$ corresponds to the upper half plane, with the real line as the boundary. In view of the decomposition (\ref{eq:psi-exp}) and (\ref{eq:chi-def}), the requirement for finiteness of the metric perturbation then implies
\begin{align}\label{eq:chi-reg}
 \lim_{{\rm Im}(w)\rightarrow 0} \cG^\ell \chi_{\ell m}&<\infty~.
\end{align}
For the solutions constructed in \cite{DHoker:2016ujz,DHoker:2016ysh,DHoker:2017mds}, the poles on the boundary $\partial\Sigma$, at which the external $(p,q)$ 5-branes of the associated 5-brane web emerge, call for special attention. We will insist that the metric fluctuation (\ref{eq:delta-ds2}) does not change the asymptotic behavior of the $AdS_6$ radius near the poles. This ensures that the poles in the perturbed solution still admit an interpretation as $(p,q)$ 5-branes. The $AdS_6$ radius in Einstein frame vanishes near the poles, $f_6^2\rightarrow 0$, as shown in sec.~3.9 of \cite{DHoker:2017mds}, while it diverges in string frame.  Demanding this behavior to remain unmodified simply amounts to requiring (\ref{eq:chi-reg}) also at the poles: as long as $\phi_{\ell m}$ is finite at the poles, the $AdS_6$ radius in the perturbed Einstein-frame metric still vanishes at the same rate as the poles are approached, due to the overall $f_6^2$ in the ansatz for the metric perturbation (\ref{eq:delta-ds2}).

For the solutions of \cite{DHoker:2017zwj} there are additional punctures in $\Sigma$, around which the supergravity fields undergo non-trivial $SL(2,\mathds{R})$ monodromy and which encode additional 7-branes in the associated 5-brane web. The Einstein-frame metric, however, is single-valued. We keep the requirement that the metric fluctuation be finite all through the interior of $\Sigma$, which ensures that the interpretation of the punctures as 7-branes is unmodified. Since $\cG$ is single-valued and also finite all through $\Sigma$ for these solutions, the punctures then do not play a special role in the fluctuation equation (\ref{eq:chi-eq}).

\subsection{Universal minimal solutions}\label{sec32}

Constructing the general solution to the partial differential equation (\ref{eq:chi-eq}) is, still, challenging.
However, with the equation in this particular form, we can immediately identify a particularly simple class of solutions, given by
\begin{align}\label{eq:min-sol}
 \chi_{\ell m}&=1~,
 &\phi_{\ell m}&=\cG^\ell~,
 & M^2&=3\ell(3\ell+5)~.
\end{align}
Since $\cG$ vanishes on all of $\partial\Sigma$ by (\ref{eq:reg2}), the regularity condition (\ref{eq:chi-reg}) excludes such constant solutions for $\ell<0$ (which would not be regular on  $S^2$ to begin with).
But for $\ell\geq 0$ these solutions satisfy the regularity conditions discussed in sec.~\ref{sec:bc}.
This special class of solutions, in fact, provides the solution with minimal $M^2$ for each $\ell$:
Multiplying eq.~(\ref{eq:chi-eq}) by $\chi_{\ell m}$ and integrating over $\Sigma$ yields
\begin{align}
 \int_\Sigma d^2w \left[\chi_{\ell m}\partial_a \left(\cG^{2\ell+2}\partial^a\chi_{\ell m}\right)+\frac{1}{6}\left(M^2-3\ell(3\ell+5)\right)\kappa^2\cG^{2\ell+1}\chi_{\ell m}^2\right]&=0~.
\end{align}
Integrating the first term by parts yields a boundary term $\propto\chi_{\ell m}\cG^{2\ell+2}\partial^a\chi_{\ell m}$, which vanishes by (\ref{eq:chi-reg}). We thus find, using that $\chi_{\ell m}$ is real,
\begin{align}
 \int_\Sigma d^2w \left[-\cG^{2\ell+2}|\partial\chi_{\ell m}|^2+\frac{1}{6}\left(M^2-3\ell(3\ell+5)\right)\kappa^2\cG^{2\ell+1}\chi_{\ell m}^2\right]&=0~.
\end{align}
Noting that $\kappa^2$ and $\cG$ are non-negative in all of $\Sigma$ and positive in the interior, we conclude that 
\begin{align}
 M^2\geq 3\ell(3\ell+5)~.
\end{align}
The solutions in (\ref{eq:min-sol}) saturate this inequality and therefore indeed produce minimal $M^2$ for a given choice of $\ell$, $m$.

\subsection{Universal non-minimal solutions}\label{sec33}

In this section we discuss a second class of solutions which is also universally present but produces non-minimal $M^2$. It is reminiscent of a second branch of solutions found for the type IIA solution in \cite{Passias:2018swc}, but differs in crucial details. As starting point we take eq.~(\ref{eq:chi-reg}), and rewrite it, noting (\ref{eq:eta}), as follows
\begin{align}\label{eq:chi-2}
 \cG\partial_w\partial_{\bar w}\chi_{\ell m}+
 (2\ell+2)(\partial_a\cG)\partial^a\chi_{\ell m}+\frac{1}{6}\left(M^2-3\ell(3\ell+5)\right)\kappa^2\chi_{\ell m}&=0~.
\end{align}
We make an ansatz for $\chi_{\ell m}$ as the difference of a holomorphic and an anti-holomorphic part,
\begin{align}
 \chi_{\ell m}(w,\bar w)&=\chi_{\ell m}^+(w)-\chi_{\ell m}^-(\bar w)~.
\end{align}
In particular, this makes $\chi_{\ell m}$ harmonic, and the first term in (\ref{eq:chi-2}) drops out. Evaluating the remaining terms more explicitly yields
\begin{align}\label{eq:chi-split}
 (\ell+1)\left((\partial_{\bar w}\cG)\partial_w\chi_{\ell m}^+-(\partial_w\cG)\partial_{\bar w}\chi_{\ell m}^-\right)
 +\frac{1}{6}\left(M^2-3\ell(3\ell+5)\right)\kappa^2(\chi^+_{\ell m}-\chi_{\ell m}^-)&=0~.
\end{align}
Using the explicit expression for $\partial_w\cG$, given by
\begin{align}
 \partial_w\cG&=(\bar\cA_+-\cA_-)\partial_w\cA_+ + (\cA_+-\bar\cA_-)\partial_w\cA_-~,
\end{align}
it is straightforward to show that eq.~(\ref{eq:chi-split}) admits a solution when $\chi_{\ell m}^+=\cA_+$ and $\chi_{\ell m}^-=\bar\cA_-$. It is given by
\begin{align}
 \chi_{\ell m}&=\cA_+ - \bar \cA_-~,
 &
 M^2&=\left(3\ell+1\right)\left(3\ell+6\right)~.
\end{align}
Note that these $\chi_{\ell m}$ are not real, and the linearized equation is solved by the real and imaginary parts separately.
They are regular in the interior of $\Sigma$ for the solutions of \cite{DHoker:2016ysh,DHoker:2017mds,DHoker:2017zwj}. The $\cA_\pm$ have logarithmic divergences at the poles on the boundary of $\Sigma$, and associated branch cuts along the boundary. The discontinuities across the branch cuts, however, cancel in the combination $\cA_+ - \bar \cA_-$, such that the $\chi_{\ell m}$ are single valued. This leaves only the logarithmic divergences at the poles. Crucially, the profile of the metric fluctuation is given by $\phi_{\ell m}$ as defined in (\ref{eq:chi-def}), not by $\chi_{\ell m}$ directly. The solutions for $\phi_{\ell m}$ are
\begin{align}\label{eq:non-min-sol}
 \phi_{\ell m}^{(1)}&=\cG^\ell\Re\left(\cA_+ - \bar \cA_-\right)~,
 &
 \phi_{\ell m}^{(2)}&=\cG^\ell\Im\left(\cA_+ - \bar \cA_-\right)~.
\end{align}
For $\ell=0$ the logarithmic singularity at the poles remains.\footnote{The complete metric perturbation (\ref{eq:delta-ds2}) in Einstein frame is finite at the poles, due to the factor $f_6^2$. But the perturbation dominates the $AdS_6$ part of the background metric and actually blows up in string frame.} For $\ell\geq 1$, however, since $\cG$ vanishes on the boundary of $\Sigma$ and at the poles (faster than logarithmically, as can be seen from the near-pole behavior derived in sec.~3.9 of \cite{DHoker:2016ujz}), these solutions are perfectly regular.

\section{Implications for the dual SCFTs}\label{sec-scft}

In this section we discuss the operators that the universal spin-2 fluctuations identified above source in the dual 5d SCFTs, and the superconformal multiplets they belong to. We also obtain the normalization of the two-point functions for these operators and discuss the relation to the five-sphere partition function.

\subsection{Superconformal multiplets}
For each of the graviton perturbations discussed in the previous section there is an operator in the spectrum of the dual SCFT, which we will call $T^{\mu\nu}_{\ell,m}$.
From the usual relation between the $AdS$ mass of a bulk field and the scaling dimension of the dual operator, $M^2=\Delta(\Delta-5)$, we conclude that the dual operators for the minimal solutions (\ref{eq:min-sol}) have scaling dimension
\begin{align}\label{eq:spin2-Delta}
 \Delta(T^{\mu\nu}_{{\rm min}, \ell,m})&=5+3\ell~.
\end{align}
To identify the multiplets that these dual operators correspond to we consult with the classification of \cite{Cordova:2016emh}\footnote{For definiteness, we use the notation and conventions of \cite{Cordova:2016emh} instead of the slightly different ones in  \cite{Buican:2016hpb} to denote the multiplets and representations.}. We first note that the Dynkin label 
$R$ in the convention of \cite{Cordova:2016emh} is always integer, rather than half integer, and related to our  $\ell$ by $R=2\ell$.\footnote{The supercharge in their eq.\ (2.56) has $R=1$, but, being a spinor, would correspond to $\ell=1/2$.} In our type IIB solution the R-symmetry is realized as the isometries of the $S^2$ and the scalar spherical harmonics on $S^2$ have  integer $\ell$, this implies that R takes even integer values. The 
fluctuation spectrum of the supergravity solution contains states of up to spin 2. The only short multiplets in table 22 of \cite{Cordova:2016emh} that fulfill this requirement are   $B_2$ and  $A_4$. The  $B_2$ representation with $R=0$ reproduces the $6d$ (gauged) supergravity multiplet. This multiplet should be present in the spectrum, as we expect our type IIB supergravity solutions to admit a truncation to $6d$ gauged supergravity. The arguments of \cite{Gauntlett:2007ma} actually suggest the stronger statement that there is a truncation which is consistent at the full non-linear level. The linear analysis is sufficient for our purposes. From the perspective of the dual SCFTs, having a $6d$ (gauged) supergravity multiplet in the spectrum simply amounts to the field theories having an energy-momentum tensor.

The natural F(4) supermultiplet candidate to host the operators $T^{\mu\nu}_{\ell,m}$ is thus $B_2$. From \cite{Cordova:2016emh}, the structure of the $B_2$ supercurrent multiplet is
\begin{align}\label{supercm}
&\boxed{[0,0]_3^{(0)}} {}^{\underrightarrow{\;\;Q\;\;}}\boxed{[1,0]_{7\over2}^{(1)}}{}^{\underrightarrow{\;\;Q\;\;}}\boxed{[0,1]_4^{(2)} \oplus [2,0]_4^{(0)}} {}^{\underrightarrow{\;\;Q\;\;}} \boxed{[1,1]_{9\over 2}^{(1)} } {}^{\underrightarrow{\;\;Q\;\;}} \boxed{[0,2]_5^{(0)}} \; .
\end{align} 
The primary in this multiplet $ [0,0]_3^{(0)}$ is  a Lorentz scalar with R-charge $R=0$ and scaling dimension $\Delta=3$. Note that  the operator dual to  the massless graviton fluctuation, $[0,2]_5^{(0)}$, arises as the $Q^4$ descendant in this multiplet.
Generalizing this structure, the minimal massive spin two fluctuations found in section \ref{sec32} with $\ell>0$ fit into  a $B_2$ multiplet where the  scalar primary $[0,0]_{3+3\ell}^{(2\ell )}$ has  scaling dimension 
\begin{align}\label{eq:B2-Delta}
\Delta_{B_2,\rm primary}= 3+ 3\ell.
\end{align}
The massive spin-2 fluctuations can  again be identified as the $Q^4$ descendant $[0,2]_{5+3\ell}^{2\ell}$. That means the operators dual to the massive spin 2 fluctuations are not primaries, but descendants in this multiplet, with the $Q^4$ precisely accounting for the difference of $2$ between the scaling dimensions  (\ref{eq:spin2-Delta}) and (\ref{eq:B2-Delta}).

The universal non-minimal fluctuation,  presented in section \ref{sec33},  is dual to  a spin-2 operator with dimension
\begin{align}\label{eq:spin2-Deltab}
\Delta(T^{\mu\nu}_{{\rm non\; min}, \ell,m})&=6+3\ell~.
\end{align}
Following the same logic as for the minimal solution we conclude that this operator is part of an $A_4$ multiplet with $R=3\ell$, where it is again a $Q^4$ descendant of a scalar primary $[0,0]_{4+3\ell}^{(2\ell)}$ with conformal dimension 
\begin{align}\label{eq:A4-Delta}
\Delta_{A_4, \rm primary}= 4+ 3\ell~.
\end{align}

\subsection{Normalization of two-point functions}

We now compute the normalization of the two-point functions of the operators $T_{\ell,m}^{\mu\nu}$ sourced by the universal graviton fluctuations of sec.~\ref{sec32} and \ref{sec33}. For the minimal solution with $\ell=0$, corresponding to the massless graviton fluctuation, the dual operator is the energy-momentum tensor, whose normalization can be fixed independently, such that the normalization of the two-point function becomes physically meaningful. This is what \cite{Chang:2017cdx} called the central charge $C_T$, which is related to the partition function on squashed spheres and therefore accessible for field theory computations using supersymmetric localization. 

Holographically, the normalization of the energy-momentum tensor two-point function can be extracted from the effective action for the $6d$ graviton.
We start from the Einstein-frame action as given e.g.\ in \cite{Polchinski:1998rr}, whose form we only need schematically,
\begin{align}\label{eq:A1}
S_{\mathrm{IIB}}&=\frac{1}{2\kappa_{10}^2} \int d^{10}x \sqrt{-g}\left(R+\dots\right)~.
\end{align}
On general grounds, expanding the action (\ref{eq:A1}) to quadratic order in a perturbation $\delta g$ yields
\begin{align}
 \delta^2 S_{\mathrm{IIB}}&=\frac{1}{\kappa_{10}^2}\int d^{10}x\sqrt{-g}\, \delta g^{MN} P_{MN}[\delta g]+\text{boundary terms}~,
\end{align}
where $P_{MN}[\delta g]$ is the equation of motion operator acting on $\delta g$, and the indices in $\delta g^{MN}$ are raised with the full $10d$ background metric. $P[\delta g]$ is given by (\ref{eq:Laplace0}), up to a factor $f_6^2$ owing to the fact that it was derived in \cite{Bachas:2011xa} for the Weyl rescaled metric where the $AdS$ factor has unit radius.
Noting also the explicit $f_6^2$ in the perturbation ansatz (\ref{eq:delta-ds2}), we have
\begin{align}
  P_{\mu\nu}[\delta g]&=f_6^2\frac{1}{\sqrt{-g}}\partial_M\sqrt{-g} g^{MN}\partial_N h_{\mu\nu}~,
  &
  \delta g^{\mu\nu}&=f_6^{-2}h^{\mu\nu}~,
\end{align}
where the indices on $h$ are now raised with the unit-radius $AdS_6$ metric. Thus, dropping boundary terms,
\begin{align}
 \delta^2 S_{\mathrm{IIB}}&=\frac{1}{\kappa_{10}^2}\int d^{10}x\,h^{\mu\nu}\partial_M\sqrt{-g} g^{MN}\partial_N h_{\mu\nu}~.
\end{align}
Using the expansion $h_{\mu\nu}=(h^{[tt]}_{\ell m})_{\mu\nu} Y_{\ell m}\phi_{\ell m}$, with normalization $\int_{S^2}Y_{\ell m}Y_{\ell^\prime m^\prime}=\Vol_{S^2}\delta_{\ell \ell^\prime}\delta_{m m^\prime}$ such that $Y_{00}=1$ and $\phi_{00}=1$ for the minimal solutions in (\ref{eq:min-sol}), yields the effective action for the $6d$ fluctuations
\begin{align}
 \delta^2 S_{\mathrm{IIB}}&=\sum_{\ell,m}C_\ell \int d^6x\sqrt{-g_{AdS_6}}\,(h^{[tt]}_{\ell,m})^{\mu\nu}\big[\square_{AdS_6}-(3\ell(3\ell+5)-2)\big] (h^{[tt]}_{\ell,m})_{\mu\nu}~.
\end{align}
The constants $C_\ell$ evaluate to
\begin{align}\label{eq:C-ell}
 C_\ell&=\frac{1}{\kappa_{10}^2}\Vol_{S^2}\int d^2w\,  4f_6^4f_2^2\rho^2 (\phi_{\ell m})^2 = \frac{8}{3\kappa_{10}^2}\Vol_{S^2}\int d^2w\, \kappa^2\cG\,(\phi_{\ell m})^2~,
\end{align}
with $dw^2=\frac{i}{2}dw\wedge d\bar w$ and where (\ref{eq:metric-combinations}) has been used to obtain the second equality. The effective 6d gravitational coupling is related to $C_0$ for the minimal $\ell=0$ solution with $\phi_{00}=1$, corresponding to the massless graviton, by $C_0=1/\kappa_6^2$.

As discussed recently in sec.~4 of \cite{Chang:2017mxc}, $C_T$ is thus directly related to $C_0$ for the minimal solution as defined above. The result (\ref{eq:C-ell}) also shows that this coefficient $C_0$, up to a universal numerical factor that is independent of the choice of solution, is precisely the finite part of the entanglement entropy for a spherical region, as computed in sec.~IV of \cite{Gutperle:2017tjo}. As verified explicitly in \cite{Gutperle:2017tjo}, the finite part of this entanglement entropy is given by the $S^5$ partition function of the dual SCFTs. This establishes a direct relation of $C_T$ to the $S^5$ partition function, and shows that the two notions of ``central charge'' are equivalent for the theories dual to the supergravity solutions discussed here.

The $C_\ell$ are in particular finite for the physically regular solutions constructed in \cite{DHoker:2016ysh,DHoker:2017mds,DHoker:2017zwj}. This follows from the discussion in \cite{Gutperle:2017tjo}, where it was shown that the integral in (\ref{eq:C-ell}) for $\ell=0$ and $\phi_{\ell m}=1$ is finite, despite the presence of poles in the differentials $\partial_w\cA_\pm$ on the boundary $\partial\Sigma$. Since the $\phi_{\ell m}$ are well behaved for the minimal solutions with $\ell\geq 0$ and the non-minimal solutions with $\ell>0$, as discussed in sec.~\ref{sec32} and \ref{sec33}, respectively, this immediately implies finiteness of the integral in (\ref{eq:C-ell}) for all these solutions. This extends to the non-minimal solution with $\ell=0$ as well, since the logarithmic singularities near the poles in $\phi_{00}$ of (\ref{eq:non-min-sol}) are mild enough for the complete integrand in (\ref{eq:C-ell}) to be integrable.

\section{Discussion}\label{sec4}

We have identified two special classes of (massive) spin-2 excitations around the warped $AdS_6\times S^2\times\Sigma$ solutions to type IIB supergravity constructed in \cite{DHoker:2017mds}. Determining the full spectrum of spin-2 fluctuations is a non-trivial problem, and the full results are expected to depend crucially on the details of the specific solution at hand. Remarkably, the fluctuations identified here are universally present, for all solutions of this form. The fluctuations are expressed directly in terms of the  basic building blocks of the background solution and satisfy natural regularity conditions and boundary conditions on $\partial\Sigma$. In the dual $5d$ SCFTs they correspond to two sets of universally present spin-2 operators, which are $Q^4$ descendants in $B_2$ and $A_4$ multiplets in the notation of \cite{Cordova:2016emh}.

These results are reminiscent of the recently obtained results on spin-2 fluctuations \cite{Passias:2018swc} around the massive type IIA solution of \cite{Brandhuber:1999np}, but differ in crucial details. The scaling dimensions (\ref{eq:spin2-Delta}) and (\ref{eq:spin2-Deltab}) superficially agree with the scaling dimensions found in \cite{Passias:2018swc}. However, as discussed in section \ref{sec32}, we only find multiplets with even integer $R$-charge, whereas  in  \cite{Passias:2018swc} even and odd integer $R$-charges appear. This can be explained by the fact that the R-symmetry is realized differently: in the type IIB solution the $R$-symmetry is realized directly as the isometries of $S^2$, while it appears as part of the isometries of a half $S^4$ in the massive IIA solutions. A second point concerns regularity of the fluctuations dual to $A_4$ multiplets, of which we find two sets. While those were found to be singular, albeit normalizable, in \cite{Passias:2018swc}, the corresponding fluctuations for the type IIB solutions identified here are regular for $\ell>0$.

The massive type IIA solution can formally be T-dualized along an $S^1$ inside the half $S^4$, to obtain a solution in type IIB. The resulting singular solution is contained as a special case in the solutions discussed here, and we have shown in app.~\ref{appa} how the PDE for the spin-2 fluctuations reduces to an ODE in that case. The implementation of regularity conditions, however, is unclear for this background and we did not attempt a general analysis. It may be interesting in that context to also study the non-Abelian T-dual.

Finally, we have studied generalizations of the notion of central charge to $5d$ SCFTs. Two objects which have been studied in that context are the coefficient $C_T$ appearing in the two-point function of the energy-momentum tensor and the five-sphere partition function. Drawing on results obtained earlier, we showed that the two notions of central charge are equivalent for the $5d$ SCFTs described by classical supergravity on the solutions of \cite{DHoker:2017mds,DHoker:2017zwj}.

We close the paper with some open questions and directions for future work:

As discussed above, the fluctuations identified here correspond to $Q^4$ descendants in short multiplets with scalar primaries. It would be interesting to identify the other members of these multiplets as well. While the full KK reduction is a technically very challenging task, the fluctuations corresponding to the remaining members of these multiplets should be related by supersymmetry transformations to the ones identified here. In particular, they should also be universal, and likely have expressions in terms of the basic building blocks of the background solutions. This should make them accessible and provide potentially interesting insights on the structure of the full KK spectrum.

An immediate question from the perspective of the dual SCFTs is whether and how the operators identified here can be realized in terms of Lagrangian fields in gauge theory deformations. The supergravity solutions of \cite{DHoker:2017mds,DHoker:2017zwj} are naturally associated to $(p,q)$ 5-brane webs, many of which have gauge theory deformations to quiver theories. It would be interesting to understand what the universality of the operators we found implies in that context and how they are realized.

\section*{Acknowledgements}

We are grateful to Constantin Bachas, Eric D'Hoker, Martin Fluder, Achilleas Passias and Paul Richmond for useful conversations. The work of MG and CFU is supported in part by the National Science Foundation under grant PHY- 16-19926. OV is supported by NSF grant PHY-1720364 and, partially, by grant FPA2015-65480-P (MINECO/FEDER UE) from the Spanish Government. MG is grateful to the International Solvay Institutes, ULB, Brussels for hospitality while this paper was finalized.

\appendix

\section{T-dual of massive IIA solution}\label{appa}
We briefly discuss the T-dual of the massive type IIA solution, as given in appendix A of \cite{Lozano:2013oma}. A choice of $\cA_\pm$ realizing this solution was given in sec.~5.6 of \cite{DHoker:2016ujz}, as
\begin{align}
 \cA_\pm&=\frac{1}{2}aw^2\mp b w~.
\end{align}
The parameters $a$, $b$ were given for $L=1$, where $L$ is the curvature radius of $AdS_6$, in \cite{DHoker:2016ujz}. For generic $L$ they are
\begin{align}
 a&=\frac{27}{16}L^4m^{1/3}~,
 &
 b&=\frac{9L^2}{8m^{1/3}}~,
\end{align}
with $m$ encoding the Romans mass. The relevant quantities for the discussion of spin-2 fluctuations are $\kappa^2$ and $\cG$, given by
\begin{align}\label{eq:kappa2-G-IIA}
 \kappa^2&=2ab(1-Z)^{1/3}~, &\cG&=\frac{ab}{3}Z~,
 &
 Z&=1-(w+\bar w)^3~.
\end{align}
With the field identifications and real coordinates discussed in \cite{DHoker:2016ujz}, defined by $\cos\theta=w+\bar w$ and $\phi_3=\frac{ia}{2bm}(w-\bar w)$, this produces the solution as given in appendix A of \cite{Lozano:2013oma}, including the factors of $L$.

\subsection{Metric perturbations}
The equation determining the spin-2 fluctuations is (\ref{eq:chi-eq}), with $\kappa^2$ and $\cG$ given in (\ref{eq:kappa2-G-IIA}). The constants $a$, $b$ drop out, and the equation becomes
\begin{align}\label{eq:chi-eq-IIA}
 Z^{-2\ell-1}\partial_a \left(Z^{2\ell+2}\partial^a\chi_{\ell m}\right)+\left(M^2-3\ell(3\ell+5)\right)\left(1-Z\right)^{1/3}\chi_{\ell m}&=0~.
\end{align}
The geometry is invariant under shifts in $w-\bar w$, and the general solution can therefore be expanded as
\begin{align}
 \chi_{\ell m}(w,\bar w)&=f(w+\bar w)e^{ik(w-\bar w)}~.
\end{align}
Eq.~(\ref{eq:chi-eq-IIA}) becomes, with a real coordinate $x=w+\bar w$ such that $Z=1-x^3$,
\begin{align}
 Z^{-2\ell-1}\partial_x \left(Z^{2\ell+2}\partial_x f\right)-k^2Zf+\left(M^2-3\ell(3\ell+5)\right)x f&=0~.
\end{align}
Setting $z=x^3$ turns this equation into 
\begin{align}
 z(1-z)\partial_z^2f-\frac{2}{3}\left((3\ell+4)z-1\right)\partial_z f+\frac{1}{9}\left(M^2-3\ell(3\ell+5)\right)f-k^2f\frac{1-z}{z^{1/3}}&=0~.
\end{align}
For $k=0$ this is a hypergeometric differential equation in standard form. The minimal solutions discussed in sec.~\ref{sec32} correspond to constant $f$ with $k=0$. The real part of the non-minimal solutions for $\chi_{\ell m}$ in sec.~\ref{sec33} corresponds to $f=z^{1/3}$ with $k=0$, while the imaginary part has non-trivial dependence on $w-\bar w$.

\end{document}